\documentclass[%
 reprint,
 amsmath,amssymb,
 aps,
]{revtex4-2}

\usepackage{graphicx}
\usepackage{dcolumn}
\usepackage{bm}
\usepackage{xcolor}
\usepackage{appendix}
\usepackage{hyperref}

\begin{document}

\preprint{APS/123-QED}

\title{Quantum real-time evolution using tensor renormalization group methods}

\author{Michael Hite}
 \email{michael-hite@uiowa.edu}
\author{Yannick Meurice}%
 \email{yannick-meurice@uiowa.edu}
\affiliation{%
 Department of Physics and Astronomy, University of Iowa\\
 30 N Dubuque St, Iowa City, IA 52242
}%


\date{\today}

\begin{abstract}
We introduce an approach for approximate real-time evolution of quantum systems using Tensor Renormalization Group (TRG) methods originally developed for imaginary time. We use Higher-Order TRG (HOTRG) to generate a coarse-grained time evolution operator for a 1+1D Transverse Ising Model with a longitudinal field. We show that it is effective and efficient in evolving Gaussian wave packets for one and two particles in the disordered phase. Near criticality behavior is more challenging in real-time. We compare our algorithm with local simulators for universal quantum computers and discuss possible benchmarking in the near future.
\end{abstract}

\maketitle


\section{Introduction}
With fully error corrected digital quantum computers expected within the next few years \cite{PhysRevX.12.021049, Egan:2021xzp, PhysRevA.107.042422}, it has become crucial to develop efficient classical computing algorithms to benchmark these machines. One of the ways we can benchmark is to real-time evolve quantum spin systems \cite{Cervera_Lierta_2018,qsimscatt, Asaduzzaman:2023wtd, Charles:2023zbl,  farreras2024simulation1dxymodel}. For 1+1D spin systems with nearest neighbor interactions, classical algorithms such as time-evolving block decimation (TEBD) with matrix product states (MPS) have been successful for approximate real-time evolution \cite{Vidal_2004, TEBD-OG-Paper, Paeckel:2019yjf}, and hybrid infinite-TEBD has allowed for extensions into 2+1D \cite{2D-TEBD}. Yet, TEBD for two and higher dimensions remains challenging and it is important to be able to compare different approaches.

For imaginary-time evolution of lattice models, coarse graining methods such as the tensor renormalization group (TRG), developed by Levin and Nave \cite{LevinNave}, have been successful in simulating 2D classical models. Attaching local tensors on sites and or links, it generates an approximate transfer matrix by keeping the long range behavior of the system, while truncating away short range contributions. TRG like algorithms have progressed into higher-dimensional models  \cite{4D-Ising, PhysRevB.103.045131, Akiyama_2020, triadnetwork, Akiyama:2024qgv, Samlodia:2024kyi, ATRG}, sigma models and gauge theories \cite{ExactBlocking, Kawauchi:2015heu,TensorReview, Akiyama:2023hvt, Bazavov:2019qih, Akiyama:2024qer}, and fermionic models \cite{GTRG-Review, Pai:2024tip, Bloch:2022vqz}. One of these algorithms is higher-order TRG (HOTRG), which allows you to choose which directions to coarse grain. Developed by Xie et al. \cite{XieHOTRG}, they applied it successfully to both two and three-dimensional classical Ising models. By setting a maximal dimensional cutoff $d_\text{cut}$, computational complexity scales like $\mathcal{O}(d_\text{cut}^{4D-1})$ and memory complexity like $\mathcal{O}(d_\text{cut}^{2D})$ for a $D$-dimensional model.

The question arises as to whether these classical coarse graining methods can be used for real-time quantum evolution. For certain quantum spin systems like the 1+1D transverse Ising model (TIM), there exists a direct mapping to a classical spin system, which in this case is an anisotropic 2D classical Ising model (CIM). Thus, one should be able to use TRG methods to simulate the dynamics of these quantum systems. To the best of our knowlwdge, the only attempt in this direction \cite{Takeda:2021mnc} has been for a $\phi^4$ theory with a Hermite polynomial expansion.

In this paper, we use the HOTRG method for real-time evolution of the TIM. In Section \ref{sec:Background} we show explicitly the mapping from the CIM to the TIM, along with the tensorial formulation of the CIM. In Section \ref{sec:Methods} we show the HOTRG algorithm, map its resultant transfer matrix to the quantum time evolution operator, and show how to prepare states and other operators in the truncated space. In Section \ref{sec:Results} we compare the eigenvalues of our coarse-grained time evolution operator to exact diagonalization, evaluate its behavior near and past criticality, and evolve Gaussian wave packets with and without a longitudinal field. We also run two-packet evolution on IBM's QISKIT quantum simulator, and discuss the respective benefits and challenges of simulation in the noisy-intermediate scale quantum (NISQ) era. Finally, in Section \ref{sec:Conclusion}, we summarize our results and discuss future applications of TRG methods for real-time quantum evolution. 

\section{Background}\label{sec:Background}
\subsection{Models}
Consider an anisotropic $N_s \times N_\tau$-site classical Ising model with only nearest neighbor interactions. The Hamiltonian for the system is given by
\begin{equation}
    H=-\sum_{x=1}^{V}\sum_{\mu=1}^{2} J_\mu \sigma_x \sigma_{x+\hat{\mu}},
    \label{eqn:classham}
\end{equation}
where the first sum is over the all the sites for a volume $V = N_s \times N_\tau$, and the second is over the coupling directions. We let $\mu = 1$ correspond to the spatial direction, and $\mu = 2$ to the temporal direction. The partition function is
\begin{equation}
    Z = \sum_{\{\sigma\}} \prod_{x} \exp\left[ \beta_s \sigma_{x} \sigma_{x+\hat{1}} \right] \exp\left[ \beta_\tau \sigma_{x} \sigma_{x+\hat{2}} \right],
    \label{eqn:partfunc}
\end{equation}
where $\beta_s = \beta J_1$ and $\beta_\tau = \beta J_2$, and the sum is over all configurations. Following Kaufman's prescription \cite{Onsager:1943jn, Kaufman, SML}, the transfer matrix method allows us to write the temporal exponential as the matrix element $\langle \sigma_{x+\hat{\tau}} | \hat{V}_1 | \sigma_x \rangle$ of the Pauli exponential
\begin{equation}
    \hat{V}_1 = \sqrt{2\sinh2\beta_\tau} e^{\Delta\tau\hat{\sigma}^x}
\end{equation}
such that $\tanh\beta_\tau = e^{-2\Delta\tau}$ and $\sinh 2\beta_\tau \sinh 2\Delta\tau = 1$, and the spatial exponential as the matrix element $\langle \sigma_{x+\hat{s}} | \hat{V}_2 | \sigma_x \rangle$ of
\begin{equation}
    \hat{V}_2 = e^{\beta_s \hat{\sigma}^z\hat{\sigma}^z}.
\end{equation}
Combining all the terms on a time slice, we get
\begin{equation}
 \hat{V}_1 = \prod_{j=1}^{N_s} \hat{V}^{(j)}_1 = \left( 2\sinh2\beta_\tau \right)^{N_s/2} e^{\Delta\tau \sum_{j=1}^{N_s} \hat{\sigma}^x_j}.
\end{equation}
and
\begin{equation}
    \hat{V}_2 = \prod_{j=1}^{N_s} \hat{V}_2^{(j)} = e^{\beta_s \hat{\sigma}^z_{j}\hat{\sigma}^z_{j+1}}.
\end{equation}
with the partition function being
\begin{equation}
    Z = \text{Tr} \left[ \left(\hat{V}_1^{1/2} \hat{V}_2 \hat{V}_1^{1/2} \right)^{N_\tau}\right].
    \label{eqn:pf-transfer-matrix}
\end{equation}
Now, the Hamiltonian for a periodic $N_s$-site 1D transverse Ising model with nearest-neighbor and transverse field interactions is
\begin{equation}
    \hat{H} = \hat{H}_\text{NN} + \hat{H}_\text{T} = -\lambda \sum_{j=1}^{N_s} \hat{\sigma}^z_{j}\hat{\sigma}^z_{j+1} - \sum_{j=1}^{N_s} \hat{\sigma}^x_j,
\end{equation}
where $\lambda$ is the nearest-neighbor coupling constant, and the coupling to the transverse field is set to one. The Suzuki-Trotter approximation of the time-evolution operator (TEO) has the form
\begin{equation}
    U(\Delta t) = e^{-i\Delta t\hat{H}_\text{T}/2} e^{-i\Delta t\hat{H}_\text{NN}} e^{-i\Delta t\hat{H}_\text{T}/2}.
    \label{eqn:TEO}
\end{equation}
Yet, under the mapping $\Delta\tau \rightarrow i\Delta t, \beta_s\rightarrow i\lambda\Delta t$, the transfer matrix in Eq. (\ref{eqn:pf-transfer-matrix}) is just the TEO in Eq. (\ref{eqn:TEO}) up to an overall constant. Thus, there is a direct mapping from the 2D CIM to the 1D TIM. We will work in the basis where the transverse field is diagonal, so that the Hamiltonian becomes
\begin{equation}
    \hat{H} = -\lambda \sum_{j=1}^{N_s} \hat{\sigma}^x_{j}\hat{\sigma}^x_{j+1} - \sum_{j=1}^{N_s} \hat{\sigma}^z_j.
\end{equation}
\subsection{Tensorial Formulation}
Referring back to the partition function in Eq. (\ref{eqn:partfunc}), we can character expand the exponential as
\begin{equation}
\begin{split}
    \exp & \left[\beta_\mu \sigma_x \sigma_{x+\hat{\mu}}\right] \\
    &= \cosh\beta_\mu\sum_{n_{x,\mu} = 0}^1 \left[\sqrt{\tanh\beta_\mu}\sigma_x \sqrt{\tanh\beta_\mu} \sigma_{x+\hat{\mu}}\right]^{n_{x,\mu}},    
\end{split}
\end{equation}
where the index $n_{x,\mu}$ is now attached to the link between the site $x$ and the neighbor site $x+\hat{\mu}$ \cite{YM-book}. Collecting all the links on a site $x$ and bringing in the sum over configurations we get
\begin{equation}
    \begin{split}
        &\sum_{\sigma_{x}=\pm1}\left(\sqrt{\tanh\beta_{s}}\sigma_{x}\right)^{n_{x-\hat{1},1}+n_{x,1}}\left(\sqrt{\tanh\beta_{\tau}}\sigma_{x}\right)^{n_{x-\hat{2},1}+n_{x,2}}\\
        &= 2\left(\sqrt{\tanh\beta_{s}}\right)^{n_{x-\hat{1},1}+n_{x,1}}\left(\sqrt{\tanh\beta_{\tau}}\right)^{n_{x-\hat{2},1}+n_{x,2}}\\
        & \qquad \times \delta\left[\left(n_{x-\hat{1},1}+n_{x,1}+n_{x-\hat{2},2}+n_{x,2}\right)\%2\right],
    \end{split}
\end{equation}
where the Kronecker delta is 1 when the sum over link indices is even, and 0 otherwise. This allows us to write Eq. (\ref{eqn:partfunc}) as
\begin{equation}
    \begin{split}
        Z = &\left(2\cosh\beta_{s}\cosh\beta_{\tau}\right)^{V} \\
        & \qquad \times \text{Tr}\left[\prod_{x=1}^{V}\hat{T}_{\left(n_{x-\hat{1},1}+n_{x,1}+n_{x-\hat{2},2}+n_{x,2}\right)}^{(x)}\right]
    \end{split}
\end{equation}
where
\begin{equation}
\begin{split}
        &\hat{T}_{\left(n_{x-\hat{1},1},n_{x,1},n_{x-\hat{2},2},n_{x,2}\right)}^{(x)}\\
        &\quad \equiv\left(\sqrt{\tanh\beta_{s}}\right)^{n_{x-\hat{1},1}+n_{x,1}}\left(\sqrt{\tanh\beta_{\tau}}\right)^{n_{x-\hat{2},1}+n_{x,2}}\\
        &\qquad\times\delta\left[\left(n_{x-\hat{1},1}+n_{x,1}+n_{x-\hat{2},2}+n_{x,2}\right)\%2\right]
\end{split}
\end{equation}
is a rank-4 tensor that sits on a site $x$ with its indices on the links (left, right, bottom, and top respectively). Now, what is important to us is not the partition function as a whole, but the tensor network. If we trace over the spatial (horizontal) indices, we recover the transfer matrix in Eq. (\ref{eqn:pf-transfer-matrix}) up to an overall constant, which means it is also proportional to the TEO in Eq. (\ref{eqn:TEO}). If we let $\tilde{U}(\beta_s, \beta_\tau)$ be the transfer matrix for an $N_s$-site model with $N_\tau$ = 1 (a single time slice), then the TEO is simply
\begin{equation}
    \hat{U}(\Delta t) = \cos^{N_s}(\lambda \Delta t) e^{i\Delta tN_s}\tilde{U}(\beta_s, \beta_\tau),
    \label{eq:TRG-TEO}
\end{equation}
where $\beta_s = i\lambda\Delta t$ and $\beta_\tau = -\ln(\Delta t)/2 - i\pi/4$. 

\section{Methods}\label{sec:Methods}
\subsection{Higher-order tensor renormalization group}
The tensor network approach to the transfer matrix allows us to use coarse graining methods like HOTRG, which approximates the tensor network by capping the number of contributing eigenstates and maintaining the dimensionality of the fundamental tensor. Since we are building a network on a single time slice, we do not perform coarse graining in the temporal direction, as the length will always be that of the fundamental tensor (length 2).

\begin{figure}
    \centering
    \includegraphics[width=1.0\linewidth]{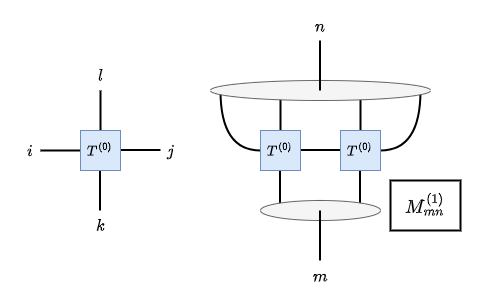}
    \caption{Tensor diagram for the fundamental tensor $T^{(0)}$ and $M^{(1)}$ used in the construction of $Q^{(1)}$.}
    \label{fig:T0-M1}
\end{figure}

\begin{figure}
    \centering
    \includegraphics[width=1\linewidth]{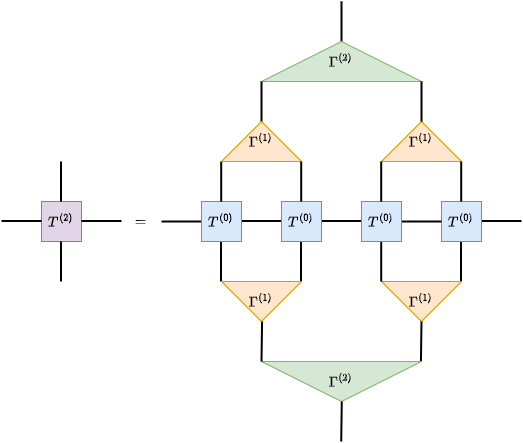}
    \caption{Tensor representation for $T^{(2)}$. The transfer matrix for a 4-site model is recovered by tracing over the spatial (horizontal) legs.}
    \label{fig:T2}
\end{figure}

Beginning with the fundamental tensor $T^{(0)}_{ijkl}$, where $(0)$ now corresponds to the zeroeth iteration of the algorithm, we seek to create another rank-4 tensor $T^{(1)}_{ijkl}$ that is equivalently $T^{(0)}_{iakm}T^{(0)}_{ajln}$. This is done by taking the spectral decomposition (SD) of a Hermitian matrix $\hat{Q}^{(1)}$ whose norm is equal to $\text{Tr}\left[ T^{(0)}T^{(0)}\right]$. As explained in Appendix \ref{app:Qmat}, at imaginary time, this is done \cite{XieHOTRG}  by using the form $Q=MM^{\dagger}$, however, at real time this leads to degenerate eigenvalues and we will rather use a $Q$ matrix designed for complex $\beta$ \cite{AlanQ}:
\begin{equation}
    \hat{Q}^{(1)} = \text{Re} \left[ \hat{M}^{(1)} \left( \hat{M}^{(1)}\right)^T\right],
\end{equation}
where $\hat{M}^{(1)}$ is the $4 \times 16$ matrix
\begin{equation}
    M^{(1)}_{mn} = M^{(1)}_{(ij)(abcd)} = T^{(0)}_{aeib}T^{(0)}_{edjc}
\end{equation}
(see Figure \ref{fig:T0-M1}). 
The SD gives
\begin{equation}
    Q^{(1)}_{ab} = \Gamma^{(1)}_{ac}\Lambda^{(1)}_{cc}\left( \Gamma^{(1)} \right)^T_{cb},
\end{equation}
where $\Lambda^{(1)}$ is the diagonal matrix of eigenvalues in hierarchical order, and $\Gamma^{(1)}$ is the corresponding matrix of eigenvectors. We can then write $\Gamma^{(1)}$ as a rank three tensor of dimensions $(2,2,4)$ to construct $T^{(1)}$:
\begin{equation}
    T^{(1)}_{ijkl} = T^{(0)}_{iabc}T^{(0)}_{ajde}\Gamma^{(1)}_{bck}\Gamma^{(1)}_{del}.
\end{equation}
The dimensions of $T^{(1)}$ are $(2,2,4,4)$, and is equivalently a two-site model. Iterating through the procedure again, $T^{(2)}$ will have dimensions $(2,2,16,16)$ for a four-site model (see Figure \ref{fig:T2}). The length of the last two indices grows like $2^{(i+1)}$, where $i$ is the iteration step. To cap this exponential growth, we introduce a cutoff in the SD step where we only keep the $d_\text{cut}$ largest contributing eigenstates of $\Gamma$. Thus, when the cutoff is reached, the remaining iterations will generate a new $T$ tensor with dimensions $(2,2,d_\text{cut},d_\text{cut}).$ We iterate $\log_2N_s$ times to generate an $N_s$-site model. The case of open boundary conditions is the matrix $T_{00ij}$ with the same proportionality constant.
\subsection{State Preparation}
\begin{figure}
    \centering
    \includegraphics[width=0.8\linewidth]{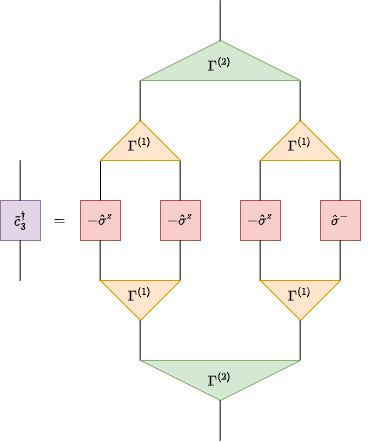}
    \caption{The transformation of $\hat{c}^\dagger_3$ from the full space to the HOTRG subspace using $\Gamma$ tensors from the time evolution operator.}
    \label{fig:hotrg-cdagger}
\end{figure}

In order to generate Gaussian wave packets, we look at the Fermionic representation of the 1D QIM, by introducing Jordan-Wigner (JW) operators at each site $j$ \cite{Sachdev, Cervera_Lierta_2018}
\begin{equation}
    \begin{split}
        \hat{c}^\dagger_j &= \left( \prod_{m=0}^{j-1} -\hat{\sigma}^z_m\right) \hat{\sigma}^-_j \\
        \hat{c}_j &= \left( \prod_{m=0}^{j-1} -\hat{\sigma}^z_m\right) \hat{\sigma}^+_j
    \end{split}
\end{equation}
(for an example of what these operators would look like in the HOTRG subspace, see Figure \ref{fig:hotrg-cdagger}). As a result of this transformation, we end up with two separate Hamiltonian's for the even and odd sectors \cite{SML}. In terms of momenta, this is expressed by having separate momenta numbers for the odd and even sectors
\begin{equation}
    k = \begin{cases} 0, \pm\frac{2\pi}{N_s}, \pm\frac{4\pi}{N_s}, \dots, \pm\frac{(N_s-2)\pi}{N_s}, \pi & \text{(odd)} \\
    \pm\frac{\pi}{N_s},\pm\frac{3\pi}{N_s},\dots,\pm\frac{(N_s-1)\pi}{N_s} & \text{(even)}\end{cases}.
\end{equation}
In position space, the Gaussian wave packet has the form
\begin{equation}
    \left| \psi_A\right\rangle = \hat{\mathcal{G}}_A\left| \Omega \right\rangle = \sum_{j=0}^{N_s-1} e^{-ik_Ax_j} e^{-r(x_j,x_A)^2/\sigma^2} \hat{c}^\dagger_{j} \left| \Omega \right\rangle,
    \label{eq:GaussWavePack}
\end{equation}
where $|\Omega\rangle$ is the vacuum, $k_A$ is the center in momentum space, $r(x_j,x_A) = |x_j-x_A|$ is the minimum distance between $x_j$ and $x_A$ in this periodic space, and $\sigma$ is the spatial width. In principle, one can build the state using the $\Gamma$-tensors alone if they know the state of each qubit, but in practice it is easier to start with the vacuum and apply operators from there.

\section{Results}\label{sec:Results}
\subsection{Energy Spectrum}
We first compare the eigenvalues of the HOTRG TEO to that of exact diagonalization (ED). The eigenvalues, up to Trotter error, have the form
\begin{equation}
    \lambda_j \simeq e^{-i\Delta t E_j},
\end{equation}
where $E_j$ is the energy of the eigenstate $\left| \lambda_j \right\rangle$. Thus, the approximate energy is
\begin{equation}
    E_j \simeq i\ln\left(\lambda_j\right)/\Delta t.
\end{equation}

\begin{figure}
    \centering
    \includegraphics[width=1\linewidth]{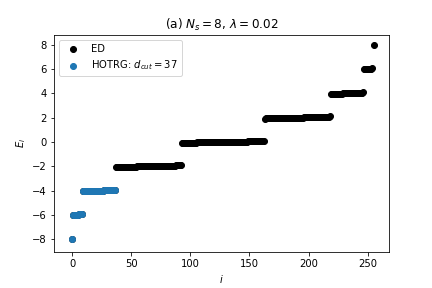}
    \includegraphics[width=1\linewidth]{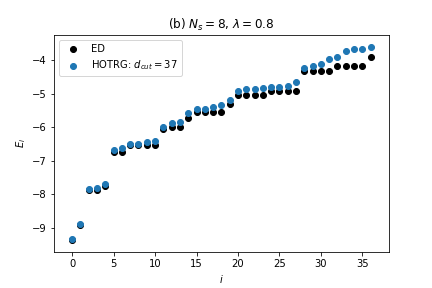}
    \caption{Energy spectrum for an 8-site TIM for (a) $\lambda=0.02$ and (b) $\lambda=0.8$. In ascending order, the terraces in (a) correspond to the vacuum and 1-8 particle sectors respectively. Plot (b) focuses in on the low energy sector.}
    \label{fig:energy-spectrum}
\end{figure}

Figure \ref{fig:energy-spectrum} (a) compares the energy spectrum for an 8-site model ($\lambda = 0.02$) of ED and the HOTRG TEO for $d_\text{cut}=37$. We see nine terraces, which are the particle sectors of the model. In ascending order, the terraces correspond to the vacuum and 1-8 particle sectors respectively. With our choice of $d_\text{cut}$, we capture the vacuum and one and two particle sectors, which shows HOTRG chooses the low energy subspace. We see excellent agreement with ED. We chose small $\lambda$ in this case to clearly distinguish the particle sectors. For larger $\lambda$, even and odd sectors mix separately, so the particle sector delineation is not as helpful. Plot (b) shows the 37 lowest energy eigenvalues for $\lambda =0.8$, with our HOTRG approximation worsening. Thus, there must be a relationship between $\lambda$ and the approximation of our HOTRG algorithm for a given system size.

\subsection{$\lambda$ Sensitivity}
\begin{figure}
    \centering
    \includegraphics[width=1.0\linewidth]{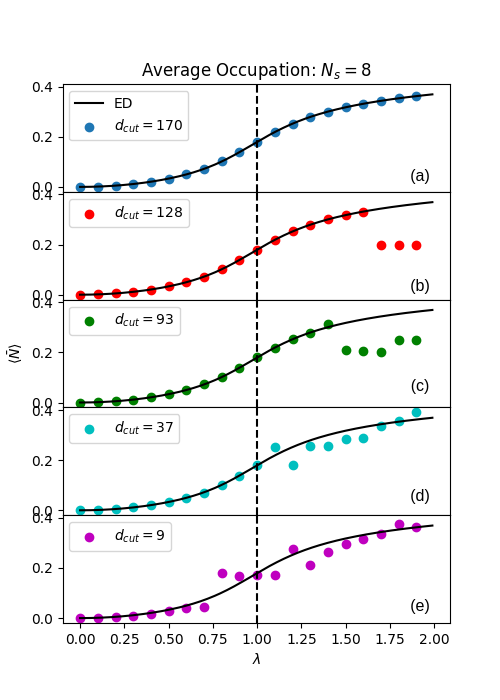}
    \caption{The average occupation of the ground state of an 8-site TIM for varying values of $\lambda$ and decreasing $d_\text{cut}$. The vertical dashed line corresponds to the critical point at $\lambda =1$.}
    \label{fig:lam-sens}
\end{figure}

Now that we have an understanding of how the HOTRG algorithm chooses states, we can consider its sensitivity to increasing $\lambda$. The two phases of the TIM are the disordered phase ($0\leq \lambda < 1$) and the ordered phase ($\lambda > 1$), with the phase transition occurring at $\lambda=1$. As with other RG schemes, HOTRG in Euclidean-time will choose long range interactions as you approach criticality, but this is not necessarily true in real-time. One way to test is to consider the average occupation of the ground state $\langle\bar{N}\rangle  = \sum_j \hat{N}_j / N_s$ for increasing $\lambda$. Figure \ref{fig:lam-sens} shows the average occupation of the 8-site ground state for $0 \leq \lambda \leq 2$ and decreasing $d_\text{cut}$. We see as $\lambda$ increases, $d_\text{cut}$ also has to increase to maintain accuracy, especially, as we go beyond criticality into the ordered phase. A possible explanation for this is the choice of basis in the TRG transformation, as in the ``particle basis'' (diagonal transverse field) \cite{qsimscatt} the disordered phase is favored. Yet, at criticality it should not matter what basis you are in, so we would see no gains by transforming into the ``spin basis'' (nondiagonal transverse field). For $N_s\geq 16$, the exponentially larger Hilbert space would make it harder to approach criticality than in the 8-site case, given the $d_\text{cut}$ dependency we observe.

\subsection{One-Particle Sector}
\begin{figure}
    \centering
    \includegraphics[width=1\linewidth]{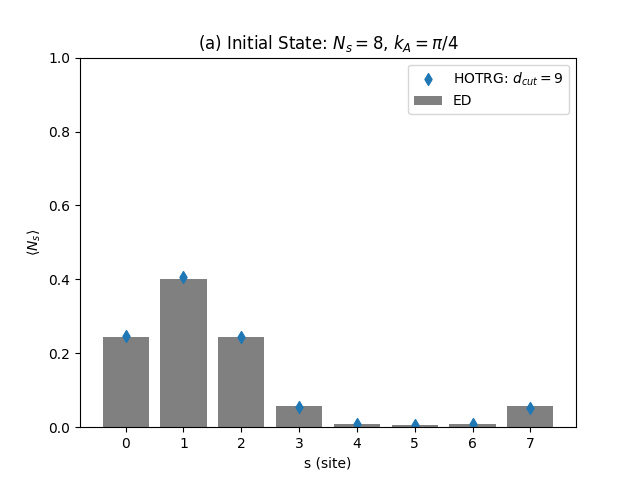}
    \includegraphics[width=1\linewidth]{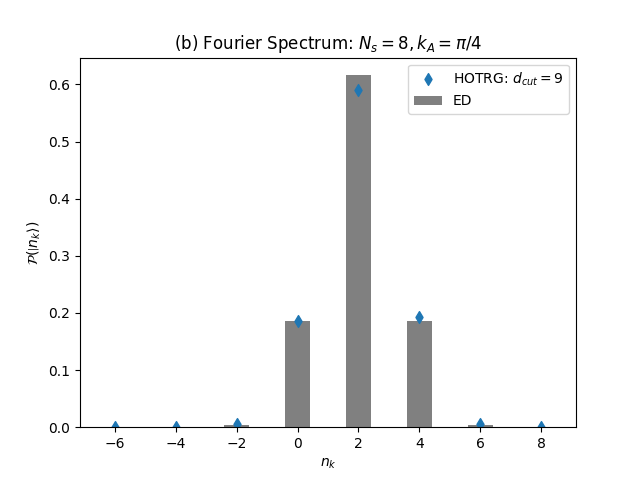}
    \caption{Our initial wave packet in (a) position and (b) momentum space with $\sigma=2$.}
    \label{fig:1-init-state}
\end{figure}

\begin{figure}
    \centering
    \includegraphics[width=1\linewidth]{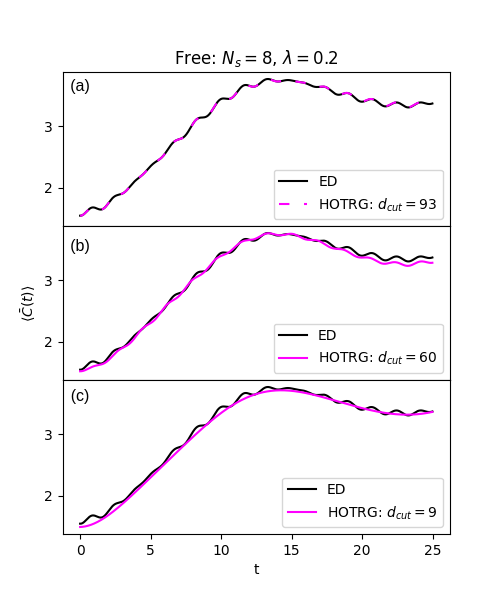}
    \caption{Evolution of $\langle \bar{C} \rangle$ of the wave packet in Figure $\ref{fig:1-init-state}$ using ED and HOTRG with decreasing $d_\text{cut}$ free of a longitudinal field.}
    \label{fig:1-part-occ}
\end{figure}

We now consider a single Gaussian-wave packet propagating in space. We initialize a $k_A = +\pi/4$ (right moving) wave packet centered at site 1 with $\sigma = 2$ (Figure \ref{fig:1-init-state} (a)). In momentum space (b), we see meaningful contributions from $k = 0, +3\pi/4$, but no other modes. For $\lambda = 0.2$ and $\Delta t = 0.01$, Figure \ref{fig:1-part-occ} shows the expectation value of the observable $\bar{C} = \sum_j j\hat{N}_j / \sum_j \hat{N}_j$, which, for open boundary conditions would give us the center of mass. We see as $d_\text{cut}$ decreases, the higher frequency, small amplitude behavior is missed, but the bulk behavior is kept.

\begin{figure}
    \centering
    \includegraphics[width=1\linewidth]{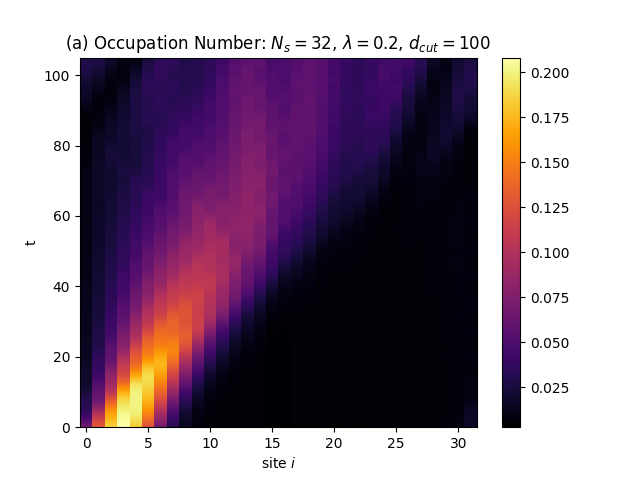}
    \includegraphics[width=1.0\linewidth]{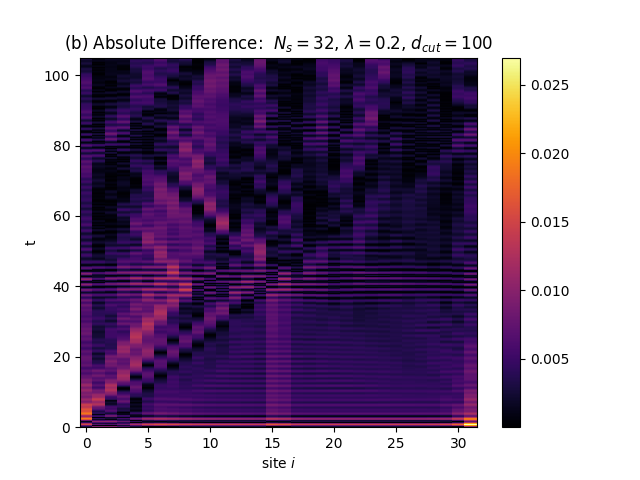}
    \caption{Evolution of a wave packet with momenta center $\pi/8$ for a 32-site TIM with $\lambda=0.2, \Delta t=0.03, d_\text{cut}=100$. Plot(a) shows the occupation number for each site and (b) the absolute difference from TEBD with max bond dimension 20.}
    \label{fig:1-part-N32}
\end{figure}

Figure \ref{fig:1-part-N32} shows the time evolution of a wave packet with momenta center $\pi/8$ for a 32-site TIM wiht $\lambda=0.2$. Plot (a) shows the occupation number of each site, and we clearly see the packet spreading out in position space as time evolves. Plot (b) shows the absolute difference at each time step for the same simulation from TEBD. All TEBD simulations were run with the ITensor Julia package \cite{ITensorArticle, ITensorCode} (See Appendix \ref{app:TEBD}). The numerical cutoff was $10^{-8}$ and max bond dimension 20. We chose absolute difference to better judge small $\langle \hat{N}_i \rangle$. The average percent difference is $4.00 \times 10^{-4}$.

With the HOTRG algorithm, we are limited by the size of the $Q$-matrix in each iteration of the algorithm, as it will have at most dimensions $(d_\text{cut}^2, d_\text{cut}^2)$, and is dense. Not only do we have to construct $Q$, but we also have to take the SVD of $Q$, which has substantial memory and time cost for large $d_{cut}$. For a PC with 16 GB of RAM and 20 GB of swap, this limits our choice of $d_\text{cut}$ to around 137 for systems with 16 or more sites. Even with access to HPC's, there would only be marginal gains for 16 sites ($d_\text{cut} \simeq 150$). For small $\lambda$ (quantum mechanics limit) we have more utility, and in the one-particle sector could even look up to 128 sites. 

\subsection{Two-particle Sector}
\begin{figure}
    \centering
    \includegraphics[width=1.0\linewidth]{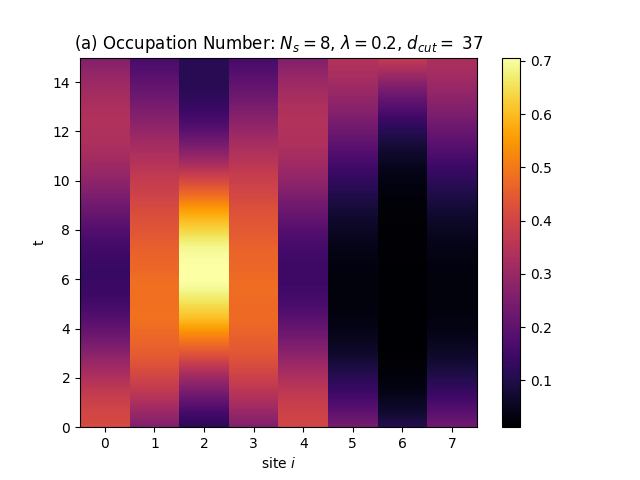}
    \includegraphics[width=1.0\linewidth]{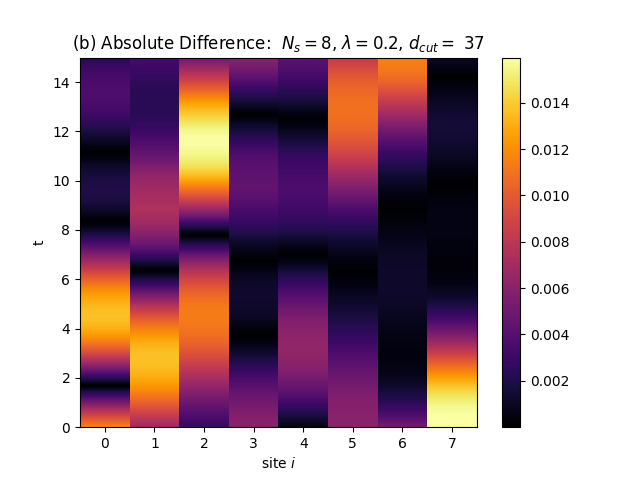}
    \caption{Occupation number (a) and absolute difference from ED (b) for a two particle state with initial position centers 0 and 4 and momenta centers $\pm 3\pi/8$ of a 8-site Ising model with $\lambda=0.2$ and $d_\text{cut}=37$.}
    \label{fig:2-part-8site}
\end{figure}

In the even sector, we switch to the half-integer momenta numbers. For 8-sites, initialize two Gaussian wave packets with momenta centers $\pm 3\pi/8$ and position centers 0 and 4. Figure \ref{fig:2-part-8site} (a) shows the occupation number $\langle \hat{N}_i \rangle$ of each site $i$, and plot (b) the absolute difference from ED. We see an average difference from ED of $4.81 \times 10^{-3}$.

\begin{figure}
    \centering
    \includegraphics[width=1\linewidth]{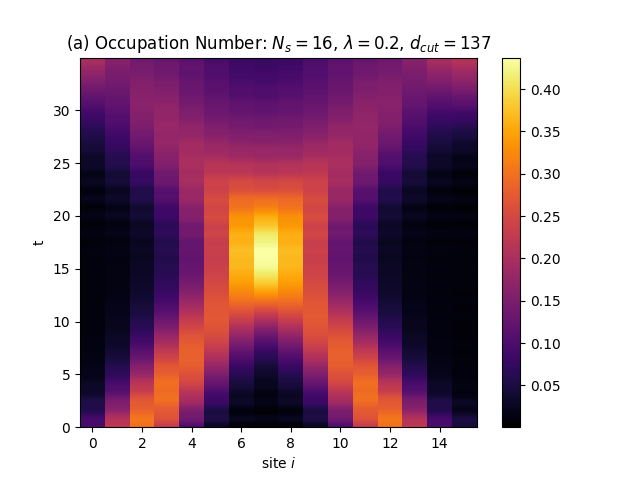}
    \includegraphics[width=1\linewidth]{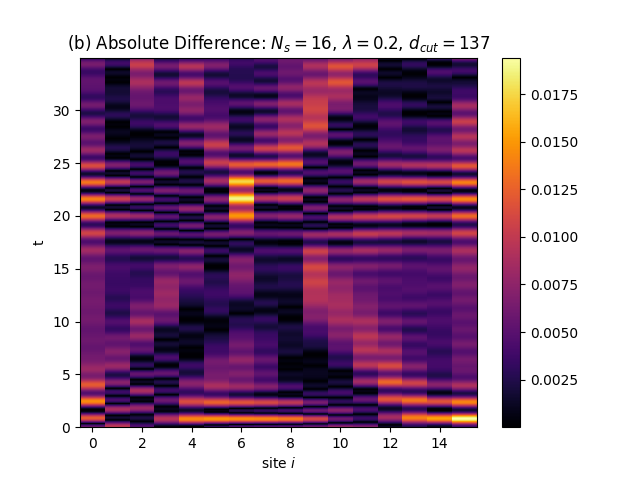}
    \caption{Occupation number (a) and absolute difference from TEBD (b) for a two particle state with initial position centers 2 and 12 and momenta centers $\pm 5\pi/16$ of a 16-site Ising model with $\lambda=0.2$ and $d_\text{cut}=137$.}
    \label{fig:2-part-16site}
\end{figure}

For 16-sites, we look at two wave packets with momenta centers $\pm 7\pi/16$ and initial positions centers 2 and 12. For $\lambda=0.2$ and $d_\text{cut}$ (vacuum and 2-particle sector), Figure \ref{fig:2-part-16site} (a) shows the propagation of the two waves with the point of maximum overlap happening at $t\approx 14$, and plot (b) the absolute difference from TEBD with default bond dimensions. For $d_\text{cut}=137$ (vacuum, one, and two-particle sector), we see an average difference from TEBD of approximately $4.74\times10^{-4}$.

\subsection{Longitudinal Perturbation}
\begin{figure}
    \centering
    \includegraphics[width=1\linewidth]{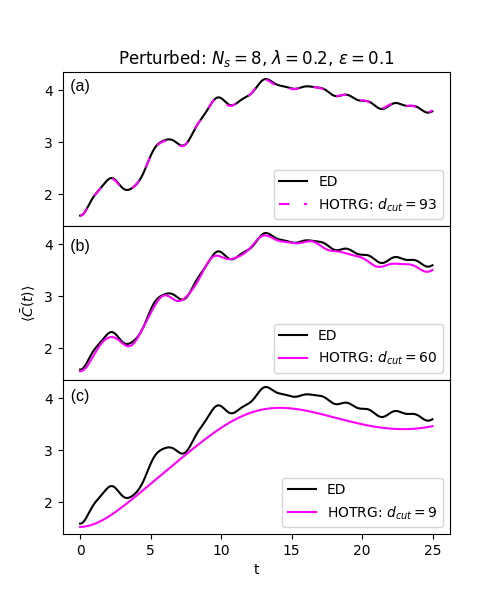}
    \caption{Evolution of $\langle \bar{C} \rangle$ of an 8-site Ising model for a single wave packet propagating through a space with a longitudinal field ($\epsilon = 0.1$).}
    \label{fig:Long-8site}
\end{figure}

\begin{figure}
    \centering
    \includegraphics[width=1\linewidth]{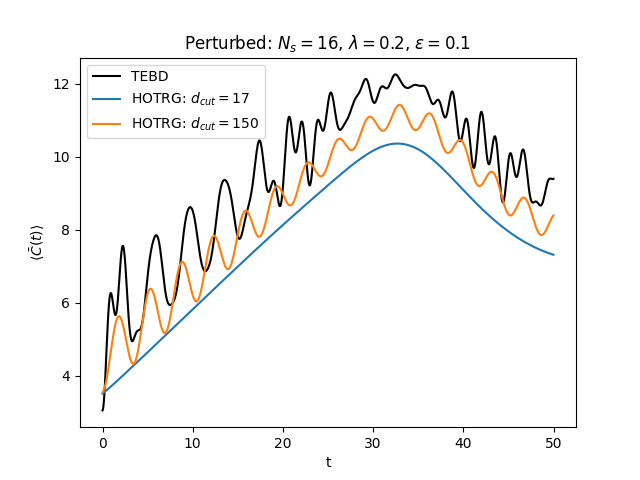}
    \caption{Expectation value of $\bar{C}$ of a 16-site Ising model for a single wave-packet through a space with longitudinal coupling $\epsilon=0.1$ and nearest-neighbor coupling $\lambda=0.2$.}
    \label{fig:Long-16site}
\end{figure}

Up to now, we have worked with an integrable theory, but we can break this integrability by introducing a longitudinal field as a perturbation. With the perturbation, the time evolution operator takes the form
\begin{equation}
    \hat{U} (\Delta t) \simeq \hat{U}^{(0)}\exp\left[-i\Delta t \epsilon \sum_{j=0}^{N_s-1} \hat{\sigma}^x_j\right],
\end{equation}
where $\hat{U}^{(0)}$ is the TEO in the free theory (see Eq. (\ref{eqn:TEO}) for the exact case and Eq. (\ref{eq:TRG-TEO}) for the HOTRG case), and $\epsilon$ is the longitudinal field coupling. In the HOTRG basis, the $\hat{\sigma}^x_j$ operators are built the same way as in Figure \ref{fig:hotrg-cdagger}, where the Pauli-$x$ matrix is applied to the $j$-th qubit, and identity to all others. As this is a perturbation, the free ground state $|\Omega\rangle$ will be used. Figure \ref{fig:Long-8site} shows the evolution of $\langle \bar{C} \rangle$ for a single wave packet moving through the longitudinal field with coupling $\epsilon=0.1$ for ED and $d_\text{cut}=93, 45$ and 9 for 8-sites. With a longitudinal field, higher frequency oscillations appear and are better accounted for with a larger $d_\text{cut}$. Figure \ref{fig:Long-16site} shows similar behavior for 16-sites, with the smaller $d_\text{cut}$ capturing none of the high-frequency oscillations due to the longitudinal field. With $d_\text{cut}=150$, we capture the higher frequency oscillations. 

\subsection{Quantum Simulation}
\begin{figure}
    \centering
    \includegraphics[width=1.0\linewidth]{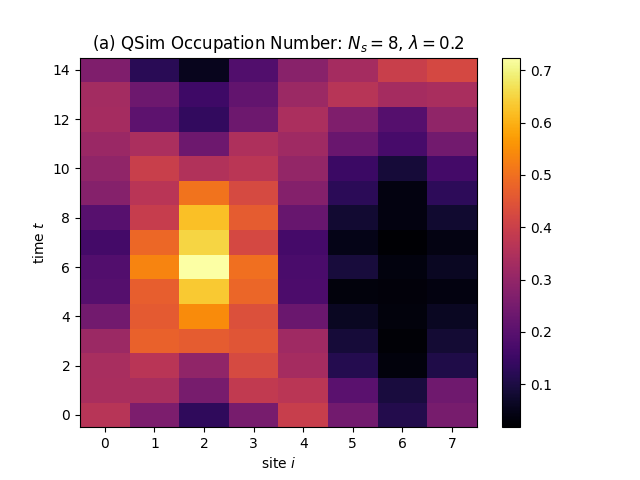}
    \includegraphics[width=1.0\linewidth]{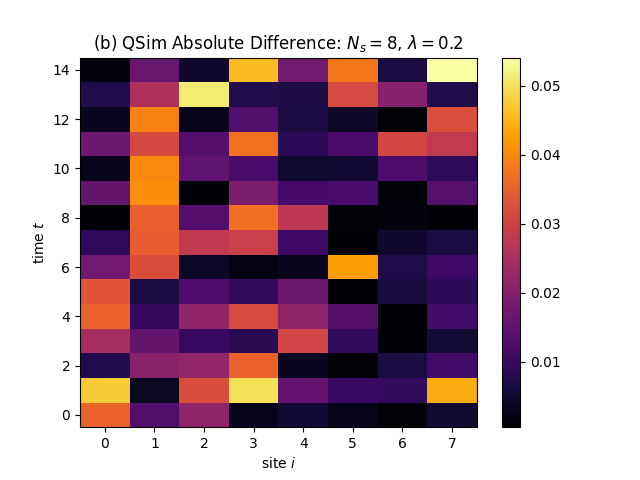}
    \includegraphics[width=1.0\linewidth]{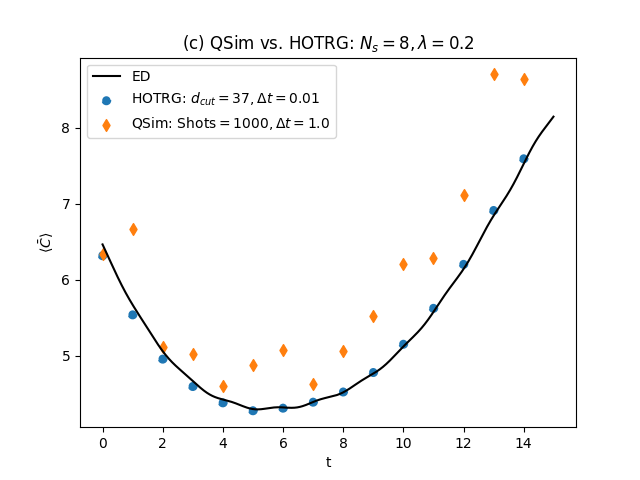}
    \caption{Quantum simulation of two particles initially centered at sites 0 and 4 and momenta centers $\pm 3\pi/8$. Plot (a) shows the occupation number of each site $i$ and (b) shows the absolute difference of the quantum simulation from ED. Plot (c) compares $\langle \bar{C}(t)\rangle$ for the same state with HOTRG for $\Delta t=0.01$.}
    \label{fig:qc-compare}
\end{figure}

We now discuss equivalent simulation on quantum computers (QC). Every method has its trade-offs, with the expensive part of the HOTRG algorithm being the creation of the time evolution operator ($Q$-matrix construction and decomposition), while time evolution and state preparation are cheap. NISQ era quantum computers are limited in all three cases, with the most expensive being state preparation. For an 8-site model, we prepare two wave packets with momenta centers $\pm 3\pi/8$, and position centers at sites 0 and 4. With Qiskit's quantum simulator \cite{QiskitCommunity2017}, we initialize the system above using the built in \verb|initialize| function with basis gates RZ, X, CNOT, and SX. The circuit depth is 1354 with 247 CNOT gates. This decomposition would not be feasible due to decoherence. Alternative methods exist for optimized wave packet preparation \cite{chai2024fermionicwavepacketscattering} and will be explored in a future publication. A single Trotter step has depth 31 and 16 CNOT gates. Figure \ref{fig:qc-compare} (a) shows the site occupation and percent error for $\lambda=0.2, \Delta t = 1.0$, 15 Trotter steps, and 500 shots on a noiseless quantum simulator, and (b) the absolute difference from Exact Trotter (ED). We chose this number of Trotter steps and shots to reflect what could be done on physical hardware. The average difference from Trotter was 0.0158.

It must be mentioned that a simulation of the same step size with HOTRG cannot be done, as the eigenstate selection for $Q$ begins to choose random high energy states rather than only low energy states. There is marginal dependence on system size and $\lambda$, but it occurs for step sizes greater than 0.05. This is why we did not do a direct comparison with the quantum simulator. But time evolution is cheap for HOTRG, so there is no benefit for us to go to larger $\Delta t$, and we can minimize Trotter error compared to today's QC's. Figure \ref{fig:qc-compare} (c) compares the evolution of $\langle \bar{C}(t)\rangle$ for our state with HOTRG for $\Delta t=0.01$. Yet, within the next few years, increased circuit depth will allow QC's to have far smaller time steps, allowing us to compare directly with our HOTRG approach.

\section{Conclusion}\label{sec:Conclusion}
We have shown that the Higher-Order Tensor Renormalization Group method, originally developed for classical lattice field theory, can be used for real-time evolution of quantum spin systems. The HOTRG algorithm selects a low energy subspace of the larger Hilbert space, and behaves quite well away from criticality in the disordered phase. As you approach criticality in real-time, HOTRG struggles to capture long-range behavior for large truncation. Using Gaussian wave packets, we can simulate one and two particle dynamics for both free and interacting theories. There have been improvements to the HOTRG algorithm, that we will explore to improve near criticality behavior \cite{Bloch:2022qyw}. More efficient TRG methods like Anisotropic TRG (ATRG) \cite{ATRG}, where computational complexity scales like $\mathcal{O}(d_\text{cut}^{2D+1})$ and memory complexity like $\mathcal{O}(d_\text{cut}^{D+1})$, will be explored. ATRG, like HOTRG, is made to extend to higher dimensions, so we could look at 2+1D Transverse Ising and others that have a classical mapping. Thus, TRG methods are a viable path forward for real-time simulation, and with future improvements, will provide a useful benchmark for quantum computers for years to come.

\section*{Acknowledgments}
This work is supported in part by the Department of Energy under Award Numbers DE-SC0019139 and DE-SC0010113. MH was also funded in part by NSF award DMR-1747426. MH would like to thank Judah Unmuth-Yockey for early discussions on TRG methods, and Muhammad Asaduzzaman and Zheyue Hang for discussions on scattering and quantum simulation. YM thanks E. Itou for discussions on real-time TRG in 2021 at a YITP virtual workshop. 

\appendix
\section{Choice of $Q$ Matrix}
\label{app:Qmat}
\begin{figure}
    \centering
    \includegraphics[width=1\linewidth]{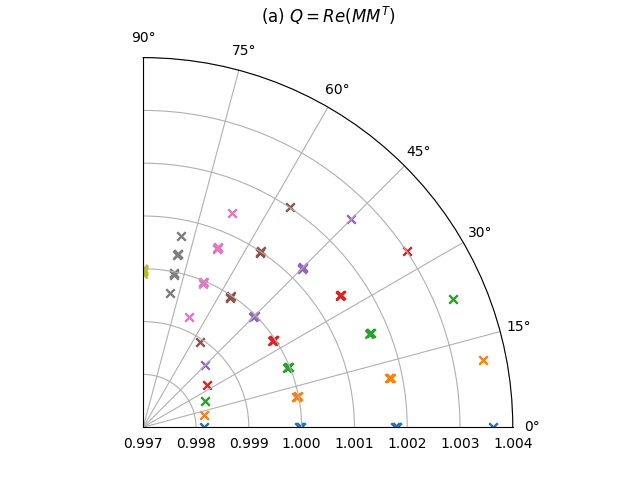}
    \includegraphics[width=1\linewidth]{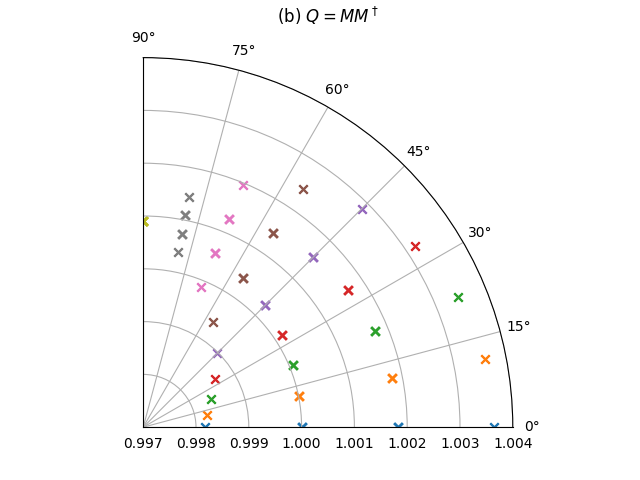}
    \includegraphics[width=1\linewidth]{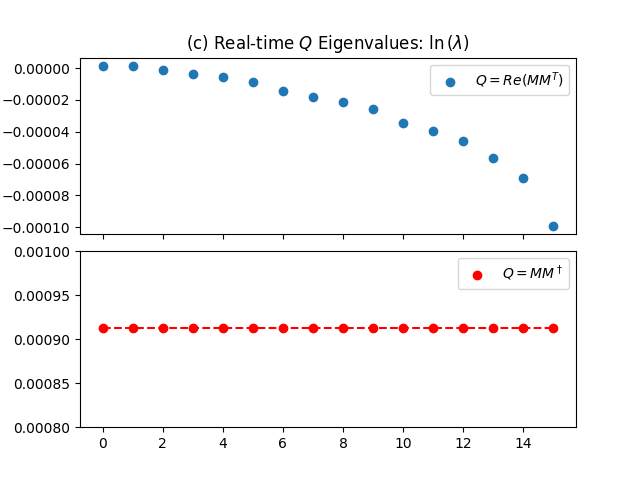}
    \caption{Eigenvalues of the $Q$-matrix as you sweep from Euclidean to real-time. Plot (a) is for the $Q$-matrix we used for real-time, and (b) what is traditionally used for Euclidean-time. Plot (c) is natural log of the eigenvalues at real-time ($\theta=\pi/2$).}
    \label{fig:Q-polar}
\end{figure}

The choice of $Q$ is dependent upon the values of $\beta_{s},\beta_{\tau}$. If $\beta_{s},\beta_{\tau}\in\mathbb{R}$ then the most straightforward choice for $Q$ is 
\begin{equation}
    Q=MM^{\dagger}.
\end{equation}
If $\beta_{s},\beta_{\tau}\in\mathbb{C}$, then the parity symmetry of $T^{(i)}$ is broken if $\Gamma$ has complex entries. Further, if $\beta_{s}$ is imaginary and $\beta_{\tau}\in\mathbb{C}$, then the eigenvalues of $Q$ are fully degenerate, which makes exclusive state selection impossible. In these cases, we can use the following choices of $Q$ where the eigenvalues are non-degenerate, and the requirements of $Q$ hold:
\begin{align}
Q	&=	\text{Re}\left[MM^{T}\right],\\
Q	&=	\text{Im}\left[MM^{\dagger}\right],\\
Q	&=	\text{Im}\left[MM^{T}\right].
\end{align}
We can show what happens to the eigenvalues of Q as we sweep from Euclidean to real-time ($\Delta\tau\longrightarrow\Delta te^{i\theta}$ for $0\leq\theta\leq\pi/2$), and compare it to the original choice $Q=MM^{\dagger}$. We have shown this to hold up to seven HOTRG iteration steps. Figure \ref{fig:Q-polar} shows the eigenvalues of $Q$ in the first iteration step as we sweep from Euclidean to real-time for (a) $Q=\text{Re}\left(MM^T\right)$ and (b) $Q=MM^\dagger$. We see that as you approach real-time the eigenvalues converge, but remain degeneration for our choice of $Q$. Plot (c) focuses in on the real-time ($\theta = \pi/2$), and we look at the natural log of the eigenvalues of each $Q$.

\section{Time Evolving Block Decimation}
\label{app:TEBD}
Time evolving block decimation (TEBD) is an efficient numerical algorithm for simulating 1D quantum systems using Matrix Product States (MPS) \cite{TEBD-OG-Paper}. We will not explain the algorithm in full, but show its application to our model. 

Our Hamiltonian contains at most single site and nearest neighbor interactions, so it can be easily written in terms of Matrix Product Operators (MPO). For Gaussian wave packets, we must do a little work, as the Jordan-Wigner operators are highly non-local. In Eq. (\ref{eq:GaussWavePack}), we created a Gaussian operator $\hat{\mathcal{G}}_A$ that we apply to the ground state. For MPO's, non-local operators like $\hat{\mathcal{G}}_A$ are very expensive, so this will not do. An alternative approach is to build the state one term at a time, so the state looks like
\begin{equation}
    \left|\psi_A \right\rangle = \sum_{j=0}^{N_s-1} \left| \phi \left(x_j,x_A\right) \right\rangle,
\end{equation}
where
\begin{equation}
    \left| \phi \left(x_j, x_A \right) \right\rangle = e^{-ik_Ax_j} e^{-r(x_j,x_A)^2/\sigma^2} \hat{c}^\dagger_{j} \left| \Omega \right\rangle,
\end{equation}
and the terms in the JW operator product are applied separately. The kets here correspond to the MPS of the state vectors above. The Gaussian wave packet is then a sum over MPS's.

\begin{figure}
    \centering
    \includegraphics[width=0.6\linewidth]{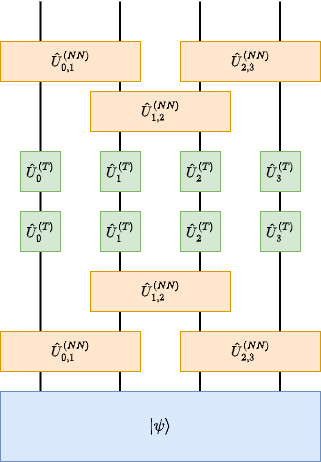}
    \caption{A single TEBD Trotter step for a 4-site Ising model with an initial MPS $\left|\psi\right\rangle$. The orange and green MPO's are defined in Eq. (\ref{eq:TEBD-ops}). The boundary term is not shown.}
    \label{fig:TEBD-Step}
\end{figure}

Now that we have state preparation, we can look at time-evolution. Let
\begin{align}
    \hat{U}^{(\text{T})}_j &= \exp\left[-i\frac{\Delta t}{2} \hat{\sigma}^z_j \right] \\
    \hat{U}^{(\text{NN})}_{j,j+1} &= \exp\left[-i\frac{\Delta t}{2} \lambda \hat{\sigma}^x_j \hat{\sigma}^x_{j+1}\right]
    \label{eq:TEBD-ops}
\end{align}
be MPO's applied to the $j$ and $j,j+1$ sites respectively. Notice we half the time step, which minimizes errors in the time evolution. For a 4-site model, a full Trotter step is given in Figure \ref{fig:TEBD-Step}.

Using the ITensor library, we can reduce computational cost by setting a cutoff on the bond dimensions of $|\psi\rangle$ and a numerical cutoff for elements of $|\psi\rangle$ \cite{ITensorArticle}. We use a numerical cutoff of $10^{-8}$ and the default dimensional cutoff unless specified.

\clearpage\bibliography{apssamp}
\end{document}